\documentclass[aps,apl,preprint,superscriptaddress]{revtex4}

\usepackage{graphicx}
\usepackage{amsmath}
\usepackage{amssymb}
\usepackage{hyperref}
\usepackage{mhchem}

\begin{document}

\title{Atomistic mechanisms governing structural stability change of zinc antimony thermoelectrics}

\author{Xiaolong Yang}
\affiliation{Frontier Institute of Science and Technology, Xi'an Jiaotong University, 710054, Xi'an, China.}

\author{Jianping Lin}
\email{jaredlin@163.com (for questions on the experimental part)}
\affiliation{School of Materials Science and Engineering, Xiamen University of Technology, Xiamen, 361024, China.}
%\affiliation{State Key Laboratory for Mechanical Behavior of Materials, Xi'an Jiaotong University, 710049, Xi'an, China}

\author{Guanjun Qiao}
\affiliation{State Key Laboratory for Mechanical Behavior of Materials, Xi'an Jiaotong University, 710049, Xi'an, China}

\author{Zhao Wang}
\email{zwangzhao@gmail.com}
\affiliation{Frontier Institute of Science and Technology, Xi'an Jiaotong University, 710054, Xi'an, China.}
\affiliation{State Key Laboratory for Mechanical Behavior of Materials, Xi'an Jiaotong University, 710049, Xi'an, China}

\begin{abstract}
The structural stability of thermoelectric materials is a subject of growing importance for their energy harvesting applications. Here we study the microscopic mechanisms governing the structural stability change of zinc antimony at its working temperature, using molecular dynamics combined with experimental measurements of the electrical and thermal conductivity. Our results show that the temperature-dependence of the thermal and electrical transport coefficients is strongly correlated with a structural transition. This is found to be associated with a relaxation process, in which a group of Zn atoms migrated between interstitial sites. This atom migration gradually leads to a stabilizing structural transition of the crystal framework, then results in a more stable crystal structure of $\beta-$\ce{Zn4Sb3} at high temperature.

\end{abstract}

\maketitle

Thermoelectric materials converting heat into electric power and vice versa, are of widespread interest for applications in energy harvesting and interconnection technologies.\cite{Bell2008} For decades most of previous studies in this field have been focusing on the strategies to improve the thermoelectric energy conversion efficiency gauged by a dimensionless figure of merit \textit{ZT}.\cite{Vineis2010}  \cite{Pei2012} \cite{Zebarjadi2012} Besides the \textit{ZT}, the thermal stability is also a key requirement for applications of thermoelectric materials,\cite{Li2010} \cite{Snyder2008} since their performance strongly depends on the material microstructure.\cite{Pei2012} \cite{Biswas2012} \cite{Singh2010} Despite its importance, only have a few attempts been made to address the thermal stability issue of thermoelectrics at their elevated working temperature.\cite{Snyder2004} \cite{Wang2013} \cite{Xu2010} \cite{Yin2011}

Zinc antimony is a typical 'phonon glass, electron crystal', one of the best thermoelectrics at moderate temperature.\cite{Caillat1997} \cite{Nylen2004} \cite{Toberer2010}In our previous experiments, it was found that $\beta-$\ce{Zn4Sb3} samples become metastable around 425 K, but above 565K it recovers its stability.\cite{Lin2014} Our previous simulation results showed that the diffusion of Zn atoms above 425K brings remarkable anharmonicity to the system vibration, and therefore is responsible for the low thermal conductivity of \ce{Zn4Sb3} in the order of those found in amorphous solids.\cite{Li2013} However, the physical reasons behind the structural stability change remained as a open question, despite this is another key to its excellent thermoelectric performance at moderate temperature.

Thus in the present work, we try to look into this issue from an atomistic point of view using molecular dynamics (MD) simulations combining with experimental measurements on the electrical and thermal conductivities of $\beta-$\ce{Zn4Sb3}. We started by building a complex hexagonal structure that corresponds to the conventional pristine structure of $\beta$-\ce{Zn4Sb3} with a concentration of Zn equal to $54.55$ at$\%$. However, this structure has been found to be thermally unstable in experiments.\cite{Izard2002} In order to obtain more realistic structure of the sample,\cite{Toberer2010} \cite{Nylen2004} \cite{Toberer2007} three Zn atoms were randomly inserted into interstitial positions at each unit cell. This yields a Zn concentration of $56.52$ at$\%$, that has been found to avoid phase separation.\cite{Rauwel2011} In our simulations we employed the classical parallel molecular dynamics package LAMMPS.\cite{PLIMPTON1995} The simulation box comprised $12\times 12\times 12$ monoclinic unit cells with 14904 atoms (containing $3888$ Sb$_{1}$ atoms, $2592$ Sb$_{2}$ atoms, $7583$ Zn$_{1}$ atoms and $841$ Zn$_{2}$ atoms), with periodic boundary conditions applied in all three spatial directions. The equations of motion were integrated using the velocity Verlet algorithm with a time step of $0.5\,\mathrm{fs}$. The simulations were performed in the isothermal-isobaric ensemble (NPT) with the Nos\'e-Hoover thermostat helping the system reach thermal equilibrium at different temperatures before any statistical analysis was performed. Atomistic interactions were described by a pairwise potential that has been successfully applied to study mechanical \cite{Li2011} and phononic behaviors of \ce{Zn4Sb3}\cite{Li2013}. Benchmark runs were performed to reproduce the characteristic $\beta$ to $\alpha$ phase transition around 250K, which manifests itself as a sharp step in the potential energy during a cooling process.

In our experiments, a group of $\beta$-\ce{Zn4Sb3} samples were prepared by plasma activated sintering system (PAS) with a starting Zn composition of $57.14$ at$\%$. The compounds were prepared from $99.999$ at$\%$-purity zinc shots and $99.99$ at$\%$-purity antimony powder(Sinpharm Chemical Reagent Co.Ltd). These elements were loaded into a quartz ampoule and sealed under vacuum down to $10^{-4}$ torr. After sealing, the ampoules were placed in a box furnace and heated to 1023 K at a rate of 375 $K/hour$ and kept at that temperature for 72 hours. Finally, they were quenched in cold water. The obtained ingots were ground in an agate mortar and sieved. Powders of upto 48$\mu$m diameter were loaded into graphite dies and sintered under vacuum ($\sim$5 Pa) using PAS at a pressure of 100 MPa and 723 K for 5 minutes. The heating rate of PAS is 400 $K/min$. The density ($\rho$) of the sintered samples were measured using the conventional Archimedes principle and were found to be $99$$\%$ of its theoretical value (6.36 $g$/$cm^{3}$).The sintered samples with a 20 $mm$ diameter were cut into several pieces for physical measurements using a spark erosion cutter. Rectangular bars with 12x3x3 $mm$ and disks with 12.7 $mm$ diameter were used for electrical transport measurements and thermal diffusivity measurements, respectively. Electrical conductivities ($\sigma$) and Seebeck coefficients were measured simultaneously by dc four-terminal method using a Linseis Seebeck Coefficient/Electric Resistance Measuring System (LSR-3) under a low-pressure helium atmosphere from room temperature to 760 K. Thermal diffusivities ($\alpha$) and specific heats ($C_{p}$) were obtained using a laser flash apparatus (NETZSCH LFA 457). The samples were coated with a thin layer of graphite to minimize errors from the emissivity of the material. The thermal conductivity ($\lambda$) was calculated from the relationship: $\lambda$=$\rho$$\alpha$$C_{p}$, where $\rho$ is the mass density.

\begin{figure}[thp]
\centerline{\includegraphics[width=13cm]{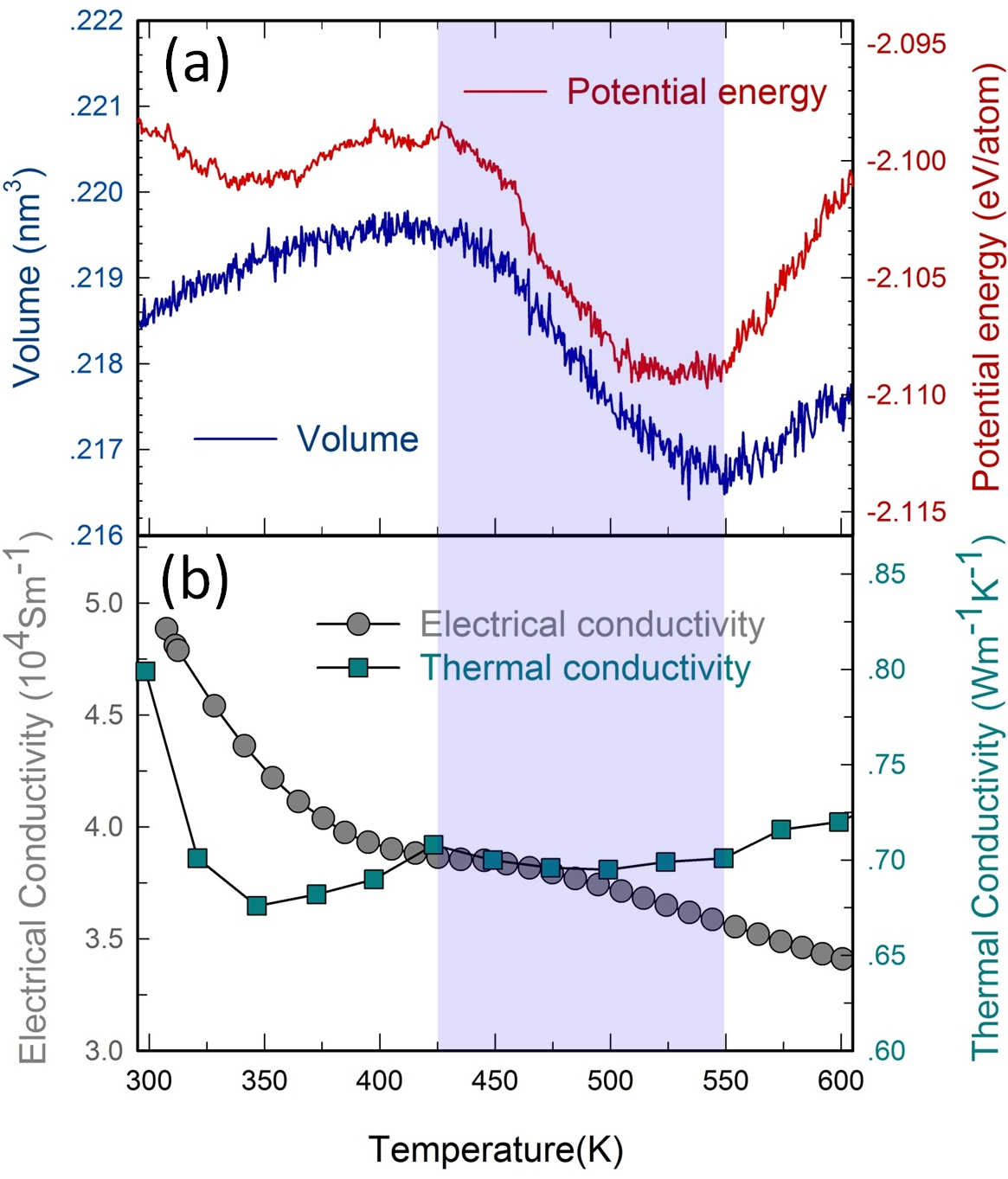}}
\caption{\label{fig1}
(a) Volume (left ordinate axis) and potential energy (right ordinate axis) variations of $\beta-$\ce{Zn4Sb3} at increasing temperature. (b) Experimentally measured electrical and thermal conductivities of $\beta-$\ce{Zn4Sb3} at increasing temperature. The highlighted background corresponds to the temperature interval at which the structure transition takes place.}
\end{figure}

In our simulation results, we obtained the crystal lattice volume and the potential energy at increasing temperature. In Fig.\ref{fig1}, we can see a volume decrease when the temperature reaches about 425K, accompanied with changes in the temperature dependence of the experimentally measured electrical and thermal conductivities. In our previous studies, we found that this volume change is triggered by a diffusion-like behavior of interstitial Zn atoms, this produces remarkable phonon anharmonicity responsible for the \ce{Zn4Sb3} low thermal conductivity.\cite{Li2013} Here we instead focus our attention on a different aspect, including a structural transition after this avalanche of atomic migration. The energy curve in Fig.\ref{fig1}(a) reveals the nature of this process, in which a group of loosely bonded atoms adapts to a different crystallographic structure in their search for a more energetically favorable arrangement. In particular, we can see that a structural transition takes place at about 550K where the relaxation process seems to be accomplished, the volume and the energy again start to increase with increasing temperature.

\begin{figure}[thp]
\centerline{\includegraphics[width=13cm]{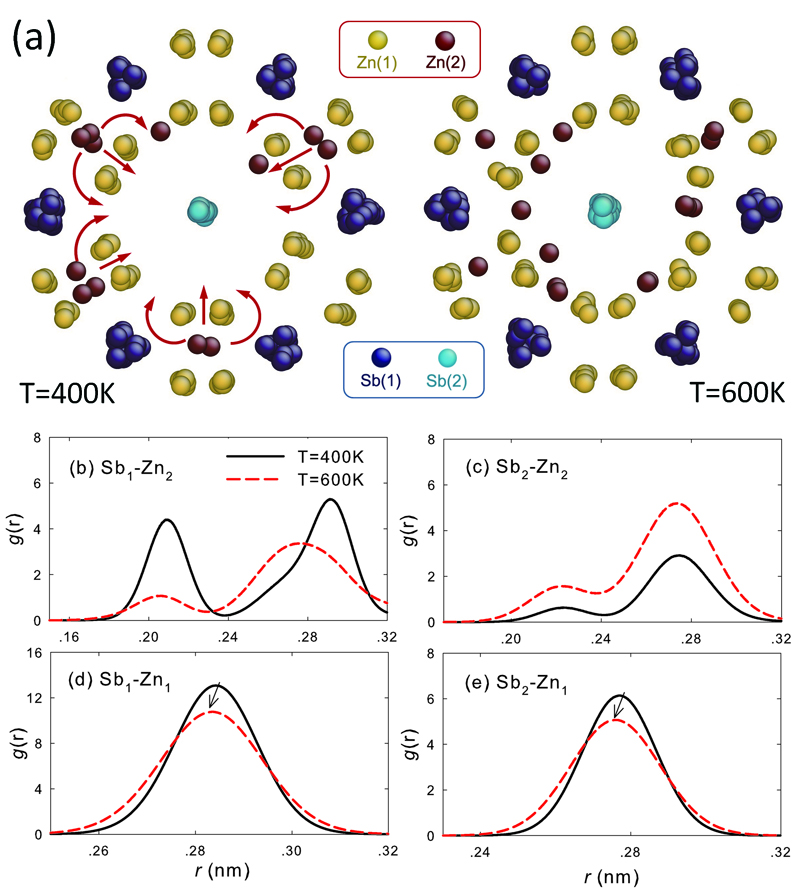}}
\caption{\label{fig2}
(a) Atomistic configurations of $\beta-$\ce{Zn4Sb3} from $\left\langle 001 \right\rangle$ before (\textit{left} at 400K) and after (\textit{right} at 600K) a structure transition. The positions of atoms are averaged over a period of 0.5 ns. Circles in four different colors show Zn atoms at crystal sites (Zn$_{1}$), Zn atoms at glass-like sites (Zn$_{2}$), and Sb atoms at two different crystal sites (Sb$_{1}$ and Sb$_{2}$), respectively. The arrows show possible displacement trajectories of Zn$_{2}$ atoms. (b-e) Radial distribution functions $g(r)$ of different pair correlations.}
\end{figure}

In order to obtain more information about this transition occurring around 550K, we have performed detailed analyses on the simulated atomistic trajectories. Fig.\ref{fig2}(a) shows two atomistic configurations of the simulated $\beta-$\ce{Zn4Sb3} before (\textit{left} panel) and after (\textit{right} panel) the structure transition. Interstitial Zn atoms (Zn$_{2}$) are found to migrate to inner sites, while Zn atoms at crystalline sites (Zn$_{1}$) seem to stay in their positions. Furthermore, we have calculated the radial distribution functions $g(r)$ of different pair correlations, through which the local relaxed structure associated with each atom is monitored. By comparing the case at 400K with that at 600K, we can see that $g(r)$ of Zn$_{2}$-Sb$_{1}$ pairs becomes lower and broader at increasing temperature [Fig.\ref{fig2}(b)], while that of Zn$_{2}$-Sb$_{2}$ pairs exhibits an inverse trend [Fig.\ref{fig2}(c)]. It is therefore double-confirmed that Zn$_{2}$ atoms  migrate from outer to inner sites and become closer to Sb$_{2}$ atoms. Regarding the main crystal lattice framework composed of Sb$_{1}$, Sb$_{2}$ and Zn$_{1}$ atoms, which are $26.087$$\%$, $17.39$$\%$ and $50.879$$\%$ occupied in our system, respectively, we can see that $g(r)$ of Sb$_{1}$-Zn$_{1}$ and Sb$_{2}$-Zn$_{1}$ pairs [Fig.\ref{fig2}(d,e)] are both broadened by increasing temperature after the structural transition, with peaks shifted to the left side. This shift indicates a change of interatomic distance in the main lattice, corresponding to the volume decrease during his structural transition.

\begin{figure}[thp]
\centerline{\includegraphics[width=13cm]{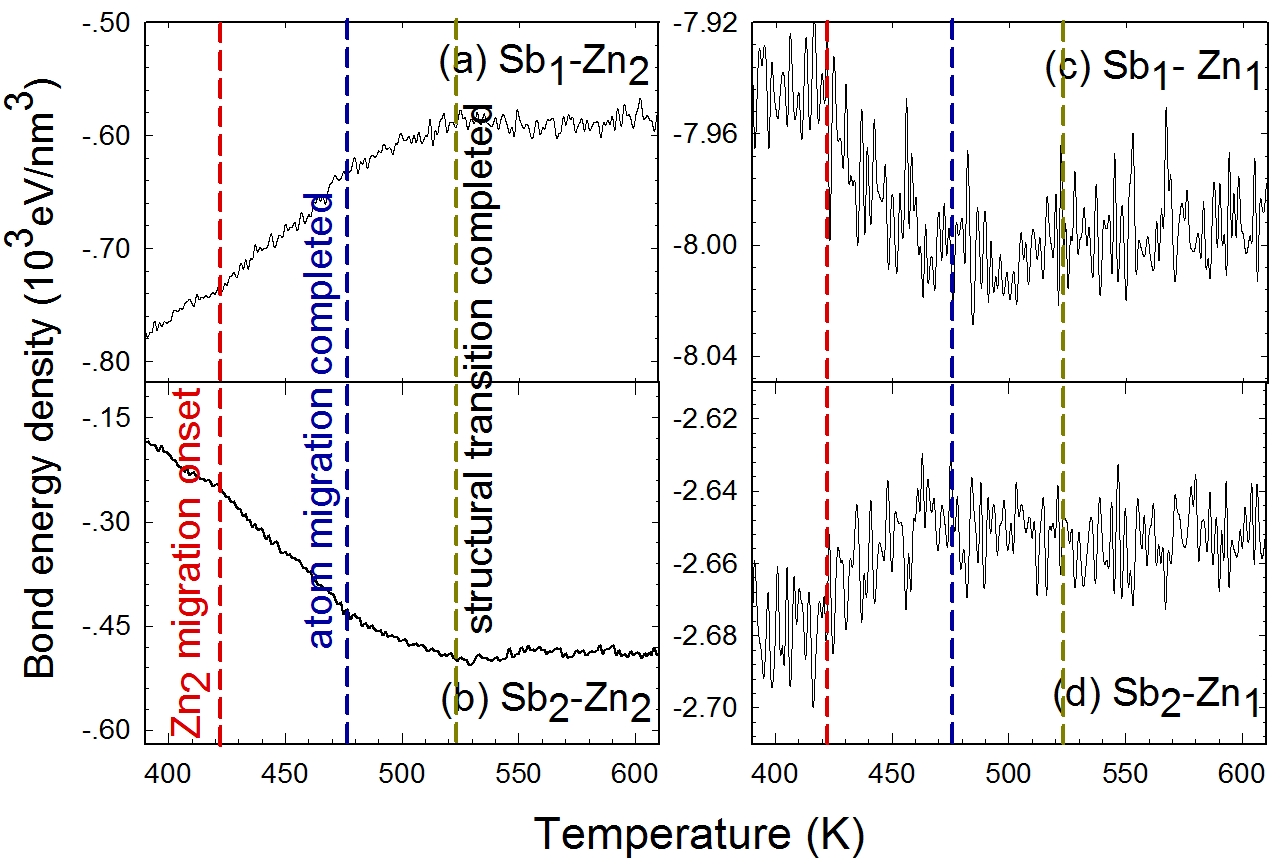}}
\caption{\label{fig3} Bond potential energy per unit cell volume between different types of atoms [(a)Sb$_{1}$ and Zn$_{2}$, (b)Sb$_{2}$ and Zn$_{2}$, (c)Sb$_{1}$ and Zn$_{1}$, (d)Sb$_{2}$ and Zn$_{1}$] \textit{versus} the temperature.}
\end{figure}

To confirm the above-described atomistic-scale details, we plot the pair potential energy between different atom sets in Fig.\ref{fig3}. It can be seen that the bonds between Zn$_{2}$ and Sb$_{1}$ atoms is weakened at increasing temperature, while those between Zn$_{2}$ and Sb$_{2}$ atoms get greatly enhanced. These are consistent with the observations commented with Fig.\ref{fig2}. The bond energy value of Sb$_{1}$-Zn$_{1}$ pair decreases while Sb$_{2}$-Zn$_{1}$ pairs increases at increasing temperature, these are also in good agreement with the above-discussed radial distribution function data.

In conclusion, we have investigated the atomistic origin of the structural stability change of $\beta-$\ce{Zn4Sb3}, which is strongly correlated with its electrical and thermal transport coefficients. Our simulation results showed that the evolution of the crystallographic stability of $\beta-$\ce{Zn4Sb3} proceeds through four steps. Firstly, thermal energy fuels a long-range relaxation process, in which a fraction of Zn$_{2}$ atoms start to migrate between interstitial sites at about 420K [Fig.\ref{fig3}(a,b)]. Secondly, the thermally-activated atom migration gradually leads to a stabilizing structural transition of the main crystal framework composed of Zn$_{1}$ and Sb [Fig.\ref{fig3}(c,d)]which are $50.879$$\%$ and $43.477$$\%$ occupied, respectively, this process seems to be accomplished at about 480K. Thirdly, the migrated Zn atoms exhibit also a local relaxation process in which they search for a more energetically favorable configuration [Fig.\ref{fig3}(a,b)]. Finally, the above-discussed processes result in a more stable crystal structure of $\beta-$\ce{Zn4Sb3} at high temperature, the system recovers it structural stability at about 520K[Fig.\ref{fig3}(a-d)]. This microscopic mechanism explains the physical reason why the thermoelectric performance of zinc antimony is sensible to slight variation in Zn composition. This is consistent with experimental observations\cite{Pedersen2010} and is applicable to super-ionic frameworks in large lattice (e.g. Sb frames) with relatively small-radius interstitials with high diffusion mobility.

We thank Dr. J. Carrete at CEA grenoble and Prof. J. Li at MIT for helpful discussions. This work is supported by a grant-in-aid of 985 Project from Xi`an Jiaotong University, the National Natural Science Foundation of China (Grant No. 11204228), the National Basic Research Program of China (2012CB619402 and 2014CB644003) and the Fundamental Research Funds for the Central Universities.

\end{document}